\documentclass[letterpaper, 10 pt, conference]{ieeeconf}  % Comment this line out
\usepackage{amsmath} % assumes amsmath package installed
\usepackage{amssymb}  % assumes amsmath package installed
\usepackage{graphicx}
\usepackage{epstopdf}
\usepackage{amsmath}
\usepackage{mathtools}
\usepackage{dsfont}
\usepackage[noadjust]{cite}

\usepackage{kantlipsum}
\usepackage{tabularx}
\usepackage{url}

\usepackage[ruled,norelsize]{algorithm2e}

\makeatletter
\newcommand{\removelatexerror}{\let\@latex@error\@gobble}
\makeatother

\newcommand{\beq}{\begin{equation}}
\newcommand{\eeq}{\end{equation}}
\newcommand{\rr}{{\mathbb R}}

\newcounter{algorithmctr}[section]
\renewcommand{\thealgorithmctr}{\thesection.\arabic{algorithmctr}}
{\refstepcounter{algorithmctr}\begin{list}{}{%
\setlength{\rightmargin}{0\linewidth}%
\setlength{\leftmargin}{.05\linewidth}
\setlength{\itemsep}{1pt}
\setlength{\parskip}{0pt}
\setlength{\parsep}{0pt}}%
\rmfamily\small
\item[]{\setlength{\parskip}{0ex}\hrulefill\par%
\nopagebreak{\bfseries\textsf{Algorithm \thealgorithmctr~}}}}%
{{\setlength{\parskip}{-1ex}\nopagebreak\par\hrulefill} \end{list}}
\IEEEoverridecommandlockouts

\title{\LARGE \bf
A decentralized algorithm for control of autonomous agents coupled by feasibility constraints
}

\author{Ugo Rosolia, Francesco Braghin, Andrew G. Alleyne, Stijn De Bruyne and Edoardo Sabbioni
\thanks{U. Rosolia is with the Department of Mechanical Engineering, University of California at Berkeley ,
Berkeley, CA 94701, USA
{\tt\footnotesize ugo.rosolia$@$ berkeley.edu}. 
}
\thanks{Francesco Braghin and Edoardo Sabbioni are with the Mechanical Engineering Department, Politecnico di Milano, Italy {\tt\footnotesize \{francesco.braghin, edoardo.sabbioni\} $@$polimi.it}. %
}
\thanks{A. G. Alleyne is with the Mechanical Science and Engineering, University of Illinois at Urbana–Champaign, Urbana, IL 61801 USA {\tt\footnotesize alleyne$@$illinois.edu}. %
}
\thanks{S. De Bruyne is with McKinsey \& Company {\tt\footnotesize stijn.debryune $@$gmail.com}. %
}}

\begin{document}

\maketitle
\thispagestyle{empty}
\pagestyle{empty}

%%%%%%%%%%%%%%%%%%%%%%%%%%%%%%%%%%%%%%%%%%%%%%%%%%%%%%%%%%%%%%%%%%%%%%%%%%%%%%%%
\begin{abstract}
 In this paper a decentralized control algorithm for systems composed of $N$ dynamically decoupled agents, coupled by feasibility constraints, is presented. The control problem is divided into $N$ optimal control sub-problems and a communication scheme is proposed to decouple computations. The derivative of the solution of each sub-problem is used to approximate the evolution of the system allowing the algorithm to decentralize and parallelize computations. The effectiveness of the proposed algorithm is shown through simulations in a cooperative driving scenario.
\end{abstract}

%%%%%%%%%%%%%%%%%%%%%%%%%%%%%%%%%%%%%%%%%%%%%%%%%%%%%%%%%%%%%%%%%%%%%%%%%%%%%%%%
\section{INTRODUCTION}
In the last decade researchers have focused on automation in several application fields and in the near future autonomous mechatronic systems will be part of our everyday lives. Under this scenario, the need for cooperative control algorithms able to manage the interactions among autonomous agents is increasing. Despite the advance in computational power allowing for solving complex tasks for a single autonomous agent in real time, it is far more challenging to control the interaction among autonomous agents \cite{1}. Indeed, when two or more agents have to interact, there could be communication limitations or the dimension of the problem could increase exponentially, and consequently the computational burden. 

In this paper we focus on dynamically decoupled systems subjected to coupling constraints. This could be the case for UAV flight formation, air traffic control, power management and several other applications \cite{1,2,3,4,5,6,7,8,9, trodden2013cooperative}. Early works in the field did not explicitly take into account the coupling constraints \cite{10,11}. For example in UAV flight control, the collision avoidance constraints are usually enforced using barrier functions, which do not guaranty safety. In \cite{8} a decentralized control strategy able to take into account hard constraints was proposed. However, the problem is solved sequentially and each decentralized optimization has to wait until the previous one is completed. Thus, for large scale systems, this approach could prove infeasible for real-time. In order to overcome this issue, the authors in \cite{9} proposed a strategy to parallelize computations. The problem is divided into $N$ sub-problems, which are solved in parallel when the agents are not coupled. In \cite{trodden2010distributed} a robust distributed MPC which allows the authors to decouple the computation is presented. A robust tube is constructed for each $i$-th agent and a local feedback controller is used to keep the agent into the tube. Therefore, communication between agents is only required to update the tube.

This work proposes a decentralized and parallelized algorithm to compute a nearly optimal solution of a specific class of non-linear non-convex problems under the assumption of no delay or loss of communication. The optimization is divided into $N$  sub-problems, similar to \cite{9}. The main contribution of this paper is to propose a communication scheme which allows for independent computation of the solution of each sub-problem. This scheme is inspired by the GMRES\textbackslash Continuation method \cite{12,13,14}, where the time evolution of a nonlinear algebraic system is traced by its derivative. The proposed scheme uses the derivative of the optimal solution to decouple the sub-problems, namely each autonomous agent approximates the behavior of the system based on its derivative. Continuations methods do not converge to the solution when the evolution of the system is discontinuous \cite{12}. Unfortunately, the coupling inequality constraints introduce a discontinuity, as shown in \cite{15}. Thus, we use a relaxed approach to deal with inequality constraints which allow us to use a continuation method. These relaxed conditions could be used also for explicit fixed time step algorithms.

This paper is organized as follow: Section II the centralized system is expressed as the summation of $N$ decentralized optimal control sub-problems. In Section III the control algorithm is described and the proposed conditions to deal with inequality constraints are derived. Section IV provides additional details on the algorithm and its range of applicability. Finally, in Section V the proposed control logic is tested on simulations of a cooperative driving scenario. Section VI provides final remarks.

%%%%%%%%%%%%%%%%%%%%%%%%%%%%%%%%%%%%%%%%%%%%%%%%%%%%%%%%%%%%%%%%%%%%%%%%%%%%%%%%
\section{PROBLEM FORMULATION}
In this section the centralized control problem is introduced. Afterwards, the relaxation method used to guarantee continuity of the optimal solution to the control problem is described. Finally, we present the decoupling strategy.
\subsection{System description}
The proposed algorithm aims to compute the trajectories of a system composed by $N$  dynamically decoupled agents. The dynamics of each agent have the following non-linear state space representation
\begin{equation}\label{eq:1}
\dot{x}_i(t) = f_i(x_i(t), u_i(t)),
\end{equation}
with $x_i \in \rr^{n_i}$ being the state vector and $u_i \in \rr^{n_{ui}}$ the control action related with the $i$-th agent. Thus, the dynamic of the overall system can be written as
\begin{equation}\label{eq:2}
\begin{aligned}
	\dot{X}(t) &= [f_1(x_1(t), u_1(t)), \cdots,f_N(x_N(t), u_N(t))]^T= \\
	&= F(X(t),U(t)),
\end{aligned}
\end{equation}
where $X(t) = [x_1(t), \cdots, x_N(t)] \in \rr^{n_1\times \cdots \times n_N}$ is the state vector and $U(t) = [u_1(t), \cdots, u_N(t)] \in \rr^{{n}_{u1}\times \cdots \times n_{uN}}$ the input vector. 

The optimal control problem consists in the minimization of $N$ decoupled cost functions over a moving time interval with fixed duration $T$:
\begin{subequations}
	\begin{align}
	&J_c^*(X(t_0)) = \inf_{U(t)} \int_{t_0}^{t_0+T} \sum_{i=1}^{N} \underbrace{h_i(x_i(t), u_i(t))}_\text{Running Cost}  dt,\\
	\text{s.t.}& \notag \\ 
	&\dot{X}(t) = F(X(t),U(t)) \\
	&I(X(t)) = [C_1(X(t)), \cdots, C_{nc}(X(t))]^T \geq 0 \label{eq:FeasConstr}
	\end{align}
\end{subequations}
where the $nc$ feasibility constraints in (\ref{eq:FeasConstr}) may couple the agents.

\subsection{The optimal control problem}
The feasibility constraints in Equation (\ref{eq:FeasConstr}) can be enforced through the cost function \cite{15,16}. Given a vector of time varying Lagrange multipliers, defined as
\begin{equation}
	\Lambda(t) = [\lambda_1(t), \cdots, \lambda_j(t), \cdots\lambda_{nc}(t)]
\end{equation}
where
\begin{equation} \label{eq:6}
	\lambda_j(t) = \begin{cases}
	\neq 0 &\mbox{If } C_j(X(t)) = 0 \\
	= 0 &\mbox{If } C_j(X(t)) > 0 \\
	\end{cases}.
\end{equation}
The centralized optimal control problem consists of the minimization of the augmented cost function
\begin{equation}
	\small {J_o(X(t),U(t)) = \int_{t}^{t+T} \underbrace{\underbrace{\sum_{i=1}^{N}h_i(x_i(t), u_i(t)}_\text{Running cost} + \underbrace{\Lambda(t)I(t)}_{\substack{\text{Enforcing feasibility} \\ \text{constraint}}} dt}_{\substack{\text{Running cost of the equivalent unconstrained problem} }}}
\end{equation}
and is defined as 
\begin{subequations}
	\begin{align}
		J_0^*(X(t_0)) &= \inf_{U(t)} \int_{t_0}^{t_0+T} J_o(X(t),U(t)) dt\\
		\text{s.t}& \\
		&\dot{X}(t) = F(X(t), U(t))
	\end{align}
\end{subequations}

\subsection{Relaxation method}
When the
inequality constraint $C_j(X(t))$ in Equation (\ref{eq:6}) is tightly satisfied after a period
where it was not, the optimal solution has a discontinuity \cite{15}. Unfortunately, continuation methods cannot be used to compute the solution at discontinuity points \cite{17}. Thus, continuation methods are not suitable to compute the optimal solution when optimality is described by the KKT conditions (Eq. (\ref{eq:6})). In order to overcome this issue, we introduce a set of slack variables to convert the inequality constraints into equality constraints,
\begin{equation} \label{eq:9}
	E(X(t)) =\begin{bmatrix}
	E_1(X(t)) \\
	\vdots \\
	E_{nc}(X(t)) \\
	\end{bmatrix} = \begin{bmatrix}
	C_1(X(t)) - z_1^2(t) \\
	\vdots \\
	C_{nc}(X(t)) - z_{nc}^2(t) \\
	\end{bmatrix} = 0.
\end{equation}

It is clear, that when the equality constraints (Eq. \ref{eq:9}) hold also the inequality feasibility constraints (Eq. (\ref{eq:FeasConstr})) are satisfied. Moreover, this problem formulation provides for the removal of the KKT conditions (Eq. \ref{eq:6}), which introduced a discontinuity. 

The relaxed optimal control problem is defined as the minimization of the cost function
\begin{equation} \label{eq:10}
	\begin{aligned}
		&J(X(t), U(t)) = \\ &=\int_{t}^{t+T} \small\underbrace{\underbrace{\sum_{i=1}^{N}h_i(x_i(t), u_i(t)}_\text{Running cost} + \underbrace{\Lambda(t)E(t)}_{\substack{\text{Enforcing feasibility} \\ \text{constraint}}} + \underbrace{\sum_{j=1}^{nc} \frac{W_z}{z_j(t)}}_{\substack{\text{Slack varible} \\ \text{effect on} \\ \text{optimality}}} dt}_{\substack{\text{Running cost of the equivalent relaxed} \\ \text{unconstrained centralized problem}}}
	\end{aligned}
\end{equation}
subject to the dynamic constraint (\ref{eq:2}). We underline that the effect of the slack variable is to add a safety margin which is determined by the tuning parameter $W_z$, and that the optimal solution of the relaxed problem does not saturates the feasibility constraints. Therefore, the solution of the relaxed problem is suboptimal for the original problem, described in Section II.B.

\subsection{Decoupling strategy}
The centralized control problem could be written as the summation of $N$ $P_i$ optimal control problems. Each $P_i$ problem is related to the $i$-th agent and is defined as,
\begin{subequations}
	\begin{align}
	J_i^*(X(t_0)) &= \inf_{u_i(t)} \int_{t_0}^{t_0+T} J_i(X(t),u_i(t))dt \\
	\text{s.t}& \\
	&\dot{x}_i(t) = f(x_i(t), u_i(t))
	\end{align}
\end{subequations}
with
\begin{equation} \label{eq:11}
\begin{aligned}
&J_i(X(t), u_i(t)) = \\ &=\int_{t}^{t+T} \small\underbrace{\underbrace{h_i(x_i(t), u_i(t)}_\text{Running cost} + \sum_{j \in A_i} [\underbrace{ \lambda_j(t)E_j(X(t))}_{\substack{\text{Enforcing feasibility} \\ \text{constraint}}} + \underbrace{ \frac{W_z}{z_j(t)}}_{\substack{\text{Slack varible} \\ \text{effect on} \\ \text{optimality}}} ] dt}_\text{Equivalent running cost of the $P_i$ problem}
\end{aligned}
\end{equation}
where $A_i$ is the set of subscripts of the inequalities $C_j(X(t))$ involving the $i$-th agent. It is clear that if at time $t$ the global optimal solutions of $N-1$ agents are known, the $P_i$ problem could be solved independently and its solution is globally optimal for the centralized relaxed problem.

\section{Algorithm}
In this section a variation to the GMRES\textbackslash Continuation methods, which allows to parallelized and decentralize computations, is presented. Moreover, we suggest a numerical strategy to handle the feasibility constraints based on their effect on optimality.

\subsection{Decentralized algorithm}

The GMRES\textbackslash Continuation method uses the derivative of the optimal solution to trace its behavior in time. For details on the numerical implementation and accuracy of continuation methods we refer to \cite{13,14,18,19,21}.

The proposed algorithm uses the derivative of the optimal solution to approximate the optimal trajectories of $N-1$ agents, enabling independent solution of each $P_i$ problem. To initialize the algorithm the derivative is computed with a centralized optimization method. After the initialization, the derivative is computed on-board on each $i$-th agent and communicated to the others. Table \ref{table:1} illustrates the algorithm’s steps. It is interesting to notice that the proposed algorithm does not introduced further numerical approximation with respect to the centralized algorithm based on continuation methods, as shown in the result section.

\begin{table}[h!] 
	\centering
	\caption{Algorithm scheme}
	\label{table:1}
	\begin{tabular}{p{1cm} p{6.9cm}}
		\multicolumn{2}{l}{
		\textbf{Initialization}} \\ \hline
		Step 1) & Compute the optimal solution and the optimal derivative with a centralized algorithm        \\
		Step 2) & Communicate the optimal solution and its derivative to all the agents   \\
		\multicolumn{2}{l}{\textbf{Iteration k}}                                                                                                            \\ \hline
		Step 3) & Each $i$-th agent integrates numerically the trajectories of the other agents \\
		Step 4) & Each $i$-th agent solves its $P_i$ problem to compute the optimal solution and its derivative \\
		Step 5) & Each $i$-th agents communicates the optimal solution and its derivative to all the other agents \\
		Step 6) & $k=k+1$ go to step 3)                                                                                                                      
	\end{tabular}
\end{table}

\subsection{Handling coupling constraints}
In this section the effect of the coupling feasibility
constraints on optimality is analyzed, and the relaxed
approach to deal with inequality constraints is introduced. As
the solution to the relaxed problem is similar to the original
one, when the feasibility inequality constraint is satisfied, the
relaxed equality constraint does not influence optimality. In
order to verify this statement, it is possible to compute the
relationship between the slack variable and the Lagrange
multiplier related to the $j$-th constraint. If $z_j(t)$ is optimal, the
following relationship for the derivative of the equivalent running cost (Eq. \ref{eq:11}) holds
\begin{equation} \label{eq:13}
	\frac{\partial (\sum_{i=1}^{N}(h(x_i(t), u_i(t) + \Lambda(t) E(t) +\sum_{j=1}^{nc} \frac{W_z}{z_j^2(t)})}{\partial z_j(t)} = 0.
\end{equation}
Combining Equation (\ref{eq:9}) and Equation (\ref{eq:13}), the explicit relation between $\lambda_j(t)$ and the state can be written as
\begin{equation} \label{eq:14}
	\lambda_j(t) = \frac{W_z}{z_j^4(t)} = \frac{W_z}{C_j^2(X(t))}.
\end{equation}
Equation (\ref{eq:14}) shows that when the inequality constraints $C_j(X(t))$ is safely satisfied, the Lagrange multiplier related with the relaxed constraint is small in magnitude. Therefore, the effect of the related relaxed feasibility constraint on optimality is negligible. When this condition occurs, we would like not to consider the unnecessary feasibility constraint, reducing the dimensions of the $i$-th optimization problem, $P_i$, in Step 4) of Table \ref{table:1}. Namely, we set a threshold value, $H_{lim}$, for which the $j$-th Lagrange multiplier of Equation (\ref{eq:10}) is set to zero
\begin{equation} \label{eq:15}
	\lambda_j(t) = \begin{cases}
	0 &\mbox{If } \lambda_j^2(t) \leq H_{lim}^2 \\
	\neq 0 &\mbox{If } \lambda_j^2(t) > H_{lim}^2 \\
	\end{cases}.
\end{equation}
Substituting in Equation (\ref{eq:15}) the relationship between the Lagrange multiplier and the system state (Eq. \ref{eq:14}), a threshold value for which the relaxed feasibility constraint has to be enforced to the problem is obtained:
\begin{equation} \label{eq:16}
\lambda_j(t) = \begin{cases}
0 &\mbox{If } C_j^2(X(t))\geq \frac{W_z}{H_{lim}} \\
\neq 0 &\mbox{If } C_j^2(X(t))< \frac{W_z}{H_{lim}} \\
\end{cases}.
\end{equation}
It is interesting to notice that these conditions (Eq. \ref{eq:16}) are similar to the KKT conditions in Equation (\ref{eq:6}), but these are suitable to apply continuation methods and fixed time step algorithms. \\

\section{Algorithm Analysis}
\subsection{Minimum principle properties}
The MGRES/Continuation method is based on the optimality conditions stated by the minimum principle \cite{14,22,23}. The minimum principle provides necessary conditions for global optimality and it is not always sufficient to compute the optimal solution \cite{16}. Therefore, it is important to analyze the algorithm to understand which class of problems could be solved with the proposed control logic.

Firstly we define the difference between weak and strong minima. Given a general optimal control problem,
\begin{subequations} \label{eq:17}
	\begin{align}
	J_g^*(x(0)) &= \inf_{u(t)} \int_{0}^{T} g(x(t), u(t)) dt \label{eq:GenCost}\\
	\text{s.t}& \\
	&\dot{x}(t) = f(x(t), u(t)).
	\end{align}
	\end{subequations}
A trajectory $x^*(t)$ is a weak minima if it minimizes the functional (Eq. (\ref{eq:GenCost})) over all the trajectories $\bar{x}(t)$ close to $x^*(t)$ in the sense of the 1-norm, meaning that 
\begin{equation}
\begin{aligned}
\small {||x^*(t) - \bar{x}(t)||_1 := \max_{0\leq t \leq T} }&\small{|x^*(t) - \bar{x}(t)| + } \\
&~~~~~~ \small{+\max_{0 \leq t \leq T} |\dot{x}^*(t) - \dot{\bar{x}}(t)| \leq \epsilon}.
\end{aligned}
\end{equation}
Conversely, a trajectory $x^*(t)$ is a strong minima if it minimizes the functional (Eq. (\ref{eq:GenCost})) over all the trajectories $\bar{x}(t)$ close to $x^*(t)$ in the sense of $0$-norm,
\begin{equation}
\small {||x^*(t) - \bar{x}(t)||_0 := \max_{0\leq t \leq T} |x^*(t) - \bar{x}(t)| \leq \epsilon}.
\end{equation}

\begin{figure}[h!]
	\centering
	\includegraphics[width=1.0\columnwidth]{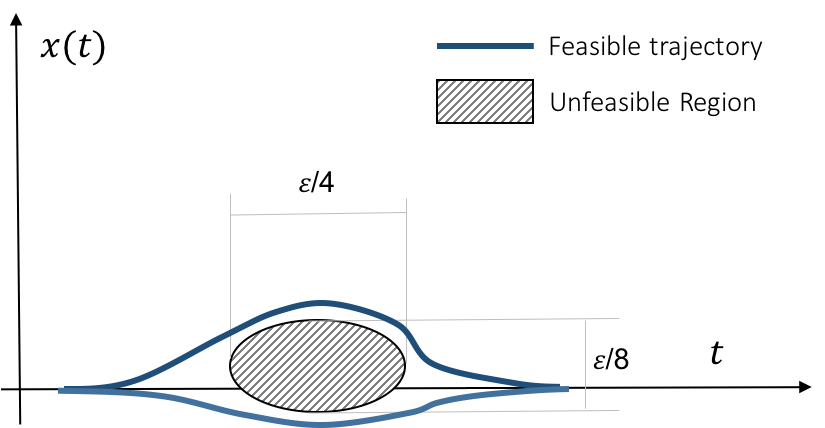}
	\caption{Domain of the feasible trajectories of the optimal control problem. The objective is to compute the trajectory $x(t)$ closest to zero which does not cross the unfeasible region.}
	\label{fig:1}
\end{figure}

The optimality conditions stated by minimum principle are satisfied for strong minima and not for weak minima \cite{16}. Therefore, the proposed algorithm is suitable to solve non- convex problems with respect to the $1$-norm, if those are convex with respect to the $0$-norm. For example, say that our control problem is to find the trajectory closest to zero, outside an unfeasible region as shown in Figure \ref{fig:1}.

From Figure \ref{fig:1} is clear that the trajectory on the left and the one the right are far in the sense of $1$-norm (the two trajectories have non-infinitesimal derivatives different in sign) and for this reason those could represent two weak minima for this problem. However, the two trajectories are close in the sense of the $0$-norm, thus the problem has just one strong minima which can be correctly computed with the optimality conditions of the minimum principle.

\subsection{Continuation method properties}
The algorithm in Section IV.C is based on a continuation method, meaning that at each time instant the optimal solution is given and the algorithm computes its derivative. This derivative is used at the next time instant to approximate the optimal solution.	

Therefore if there are more trajectories satisfying the minimum principle, the algorithm would compute the evolution in time of the given trajectory. However, there could be issues at bifurcation points where the optimal solution has two possible derivatives. This particular situation is discussed in the next section.

\subsection{Non-convex problem}
\begin{figure}[h!]
	\centering
	\includegraphics[width=1.0\columnwidth]{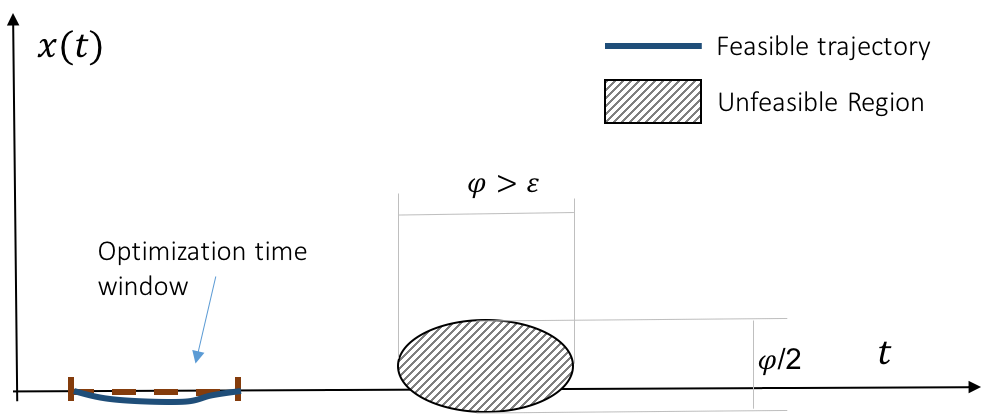}
	\caption{ Optimal control problem on a time moving window. The objective is to compute the trajectory closest to zero, outside the unfeasible region.}
	\label{fig:3}
\end{figure}

Combining the properties of the minimum principle and the continuation methods we are able to solve a particular type of non-linear non-convex problem. Indeed the algorithm is able to take non-convex decisions if the candidate trajectories are close in the sense of the 0-norm. This property has a key importance in control problems where the optimization is performed on a moving time interval. In Figure \ref{fig:3} the domain of an optimal control problem similar to the one in Section IV.A is shown. Here the objective is to compute, on a moving time interval, the feasible trajectory closest to zero.

\begin{figure}[h!]
	\centering
	\includegraphics[width=1.0\columnwidth]{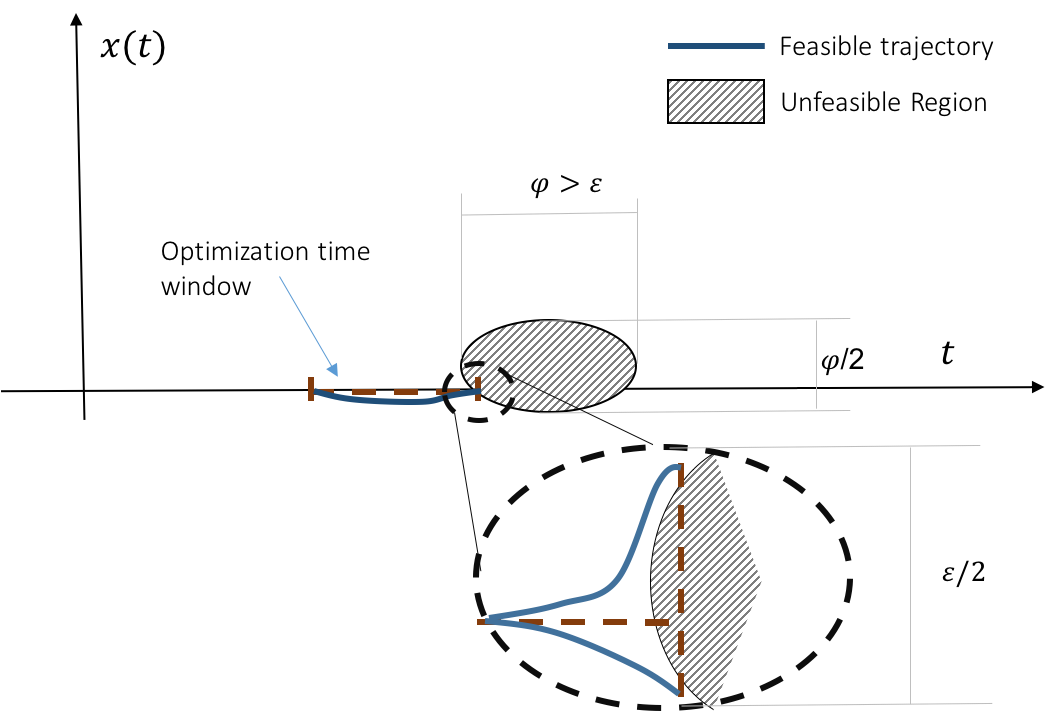}
	\caption{Representation of a bifurcation point. As soon as the unfeasible region enters the optimization window the there are two weak local minima.}
	\label{fig:4}
\end{figure}

In Figure \ref{fig:3} the unfeasible region is outside the optimization window, thus the problem is convex. When the optimization windows moves in time, as soon as it encounters the unfeasible region, the problem becomes non-convex. Indeed there are two weak minima as shown in Figure \ref{fig:4}.


Figure \ref{fig:4} depicts bifurcation point mentioned in Section IV.B. As soon as the unfeasible region enters the optimization windows, the global optimal solution bifurcates into two local optimal solutions with respect to the $1$-norm. However, as shown in Figure \ref{fig:4}, the two trajectories are close with respect to the $0$-norm, and the optimality conditions of the minimum principle allow to compute the optimal derivative for the unique strong minima. Therefore, the algorithm correctly choses the global optimal trajectory.

\begin{figure}[h!]
	\centering
	\includegraphics[width=1.0\columnwidth]{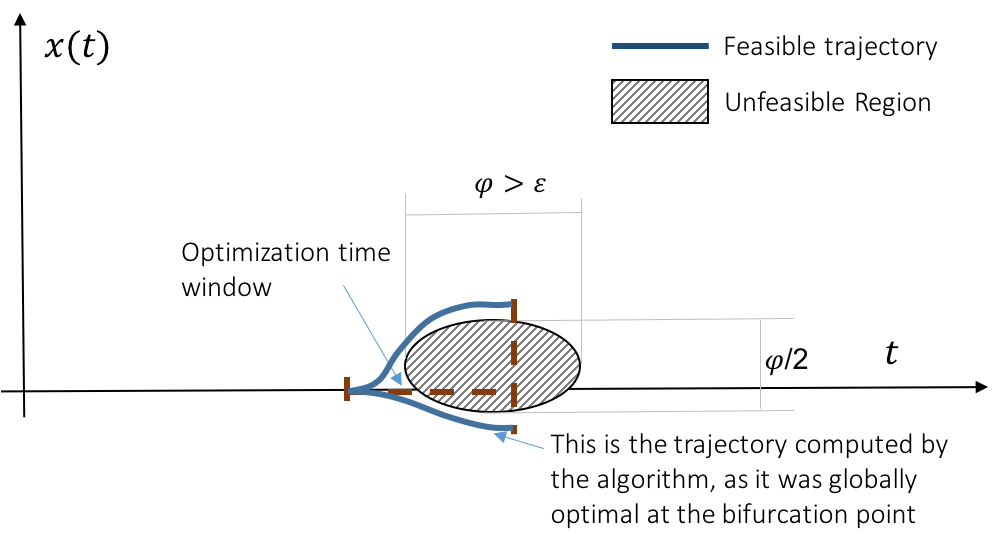}
	\caption{ Evolution of the optimal solution after the unfeasible region has entered the optimization window.}
	\label{fig:5}
\end{figure}

Afterwards, when the unfeasible region is almost, completely inside the optimization window, the continuation algorithm follows the trajectory which was globally optimal at the bifurcation point (Fig. \ref{fig:5}).

\section{Results}
The algorithm is tested on a cooperative driving scenario, where autonomous vehicles are driving on the same roadway at different target speeds. In particular, the algorithm is used to compute the collision free-trajectories of each autonomous vehicles. The vehicles are modeled with a simplified system; this choice for the trajectory planning phase is well-established in literature \cite{24,25,26}. It is important to note that this problem is suitable to test our algorithm as each vehicle has to take a non-convex decision during overtaking maneuvers. Moreover, we assume that no safety maneuvers are needed to guaranty the existence of the derivative required in Section III.A. 

Simulation was performed on a Windows computer featuring an Intel CORE i5 processor using Matlab 2013b. In order to measure the computational time, a stand-along executable mex-function has been compiled for each agent. This function could be used on Linux PCs and experimental results are envisaged for the future.

\subsection{Comparison between decentralized and centralized approach}
The agents in section III.A represent autonomous vehicles and are modeled using a Single Point Mass Model in a curvilinear abscissa reference frame, for more details \cite{27, 28}. The cost function of each vehicle is designed for lane keeping at a cruise velocity:
\begin{equation}
\begin{aligned}
	h_i(x_i(t), u_i(t) = W_{1i}(y_1 - &y_{i_{target}})^2 + W_{2i} \dot{y}_i^2 \\
	&+W_{3i}(\dot{s}_i - V_{i_{target}})^2 + W_{4i} \dot{\theta}_i^2
\end{aligned}
\end{equation}
where $s_i$ represents the distance traveled along the roadway mid-line, and $y_i$ the lateral distance between the vehicle’s center of gravity and the roadway mid-line. The inputs, $V_i$ and $\theta_i$, are the velocity and the heading angle, respectively. $W_{ji}, ~ \forall j \in [1, \cdots, 4]$ are the weighting parameters. More details on this curvilinear reference frame are given in \cite{27}, \cite{28} and \cite{29}.

Finally, the feasibility constraints in Section II.C are expressed as ellipses
\begin{equation}
	C(s_i,y_i,s_j,y_j) = \frac{(s_i-s_j)^2}{2l} +\frac{(y_i-y_j)^2}{2w} -1
\end{equation}
where the axes are chosen accordingly with vehicle length, $l$, and width, $w$: 4 and 2 meters, respectively. In this example, for each problem $P_i$ the distance between the $i$-th agent and the $j$-th agent is given by $z_j(t)$. From Equation (\ref{eq:14}) and form our choice of $W_z = 7$, when $z_j(t)= 60m$  then $\lambda_j(t) \simeq 0.5 \cdot 10^{-7} $. Therefore we picked the threshold $H_{lim} = 0.5 \cdot 10^{-7}$ so that, when the relative distance between two agents is greater that $60m$, $\lambda_j(t)$ is set to zero and the agents are decoupled. Note that $60m$ is the threshold distance used in commercial blind spot detection system. 

\subsection{Simulation Results}
\subsubsection{Comparison with a centralized algorithm}
In this section two simulations with the same boundary conditions are carried out. The first one uses the proposed decentralized algorithm and the second one uses a centralized GMRES\textbackslash Continuation algorithm. The solutions are compared to test optimality, as the solution computed with the centralized method is optimal for the relaxed problem.

\begin{figure}[h!]
		\centering
		\includegraphics[width=1.0\columnwidth]{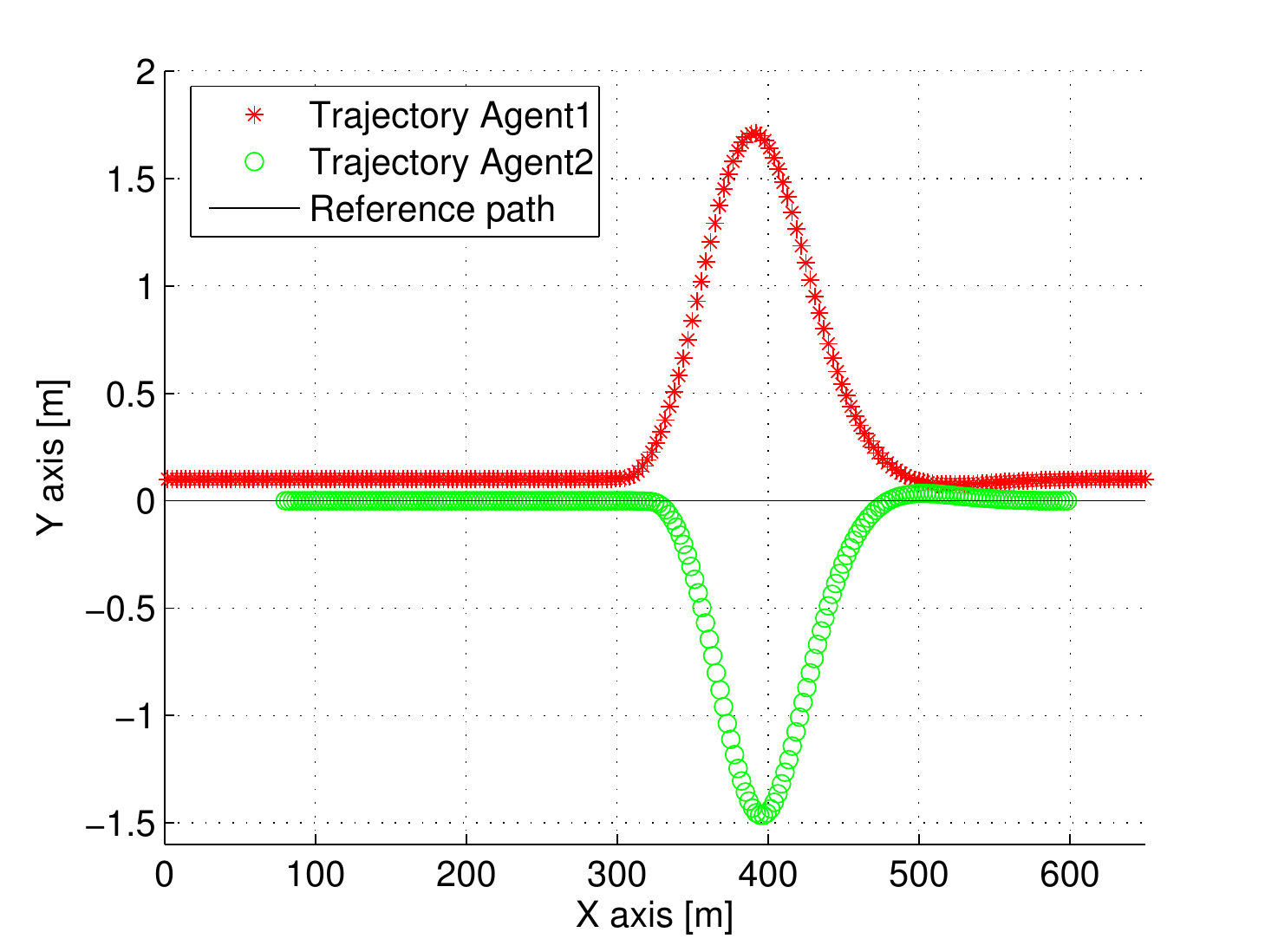}
		\caption{Trajectories of two agents traveling on the same target path at different target speed: $30 m/s$ the agent in red, $24 m/s$ the agent in green.}
		\label{Res1:0}
	\end{figure}

As shown in Figure \ref{Res1:0} two agents are traveling on the same straight path at different target velocities; therefore the faster agent overtakes the slower one. During the overtaking maneuver the agents move sideways from the centerline, so that the overall derivative of the steering angle and lateral velocity are minimized. Coefficients and boundary conditions used in the simulation for the two agents ($i=1$ and $i=2$) can be found in Tables II and III. 

\begin{table}[h!]
	\centering
	\caption{Simulation Coefficients}
	\label{my-label}
	\begin{tabular}{llllllll}
		$W_{1i}$ & $W_{2i}$ & $W_{3i}$ & $W_{4i}$ & $W_{z}$ & $T$ & $\Delta T$ & $H_{lim}$     \\     \hline        
		$0.55$      & $0.05$      & $9$         & $145$       & $7$        & $2$ & $20$       & $5 \cdot 10^{-7}$
	\end{tabular}
\end{table}

\begin{table}[h!]
	\centering
	\caption{Agents's target velocity and lateral offset}
	\label{my-label}
	\begin{tabular}{lllllll}
		Agent                & $i=1$ & $i=2$ & $i=3$ & $i=4$  & $i=5$  & {[}units{]} \\ \hline
		$y_{i_{target}}$ & $0.1$ & $0$   & $0$   & $-0.1$ & $-0.1$ & {[}m{]}     \\
		$V_{i_{target}}$ & $30$  & $24$  & $24$  & $18$   & $18$   & {[}m/s{]}  
	\end{tabular}
\end{table}

\begin{figure}[h!]
	\centering
	\includegraphics[width=1.0\columnwidth]{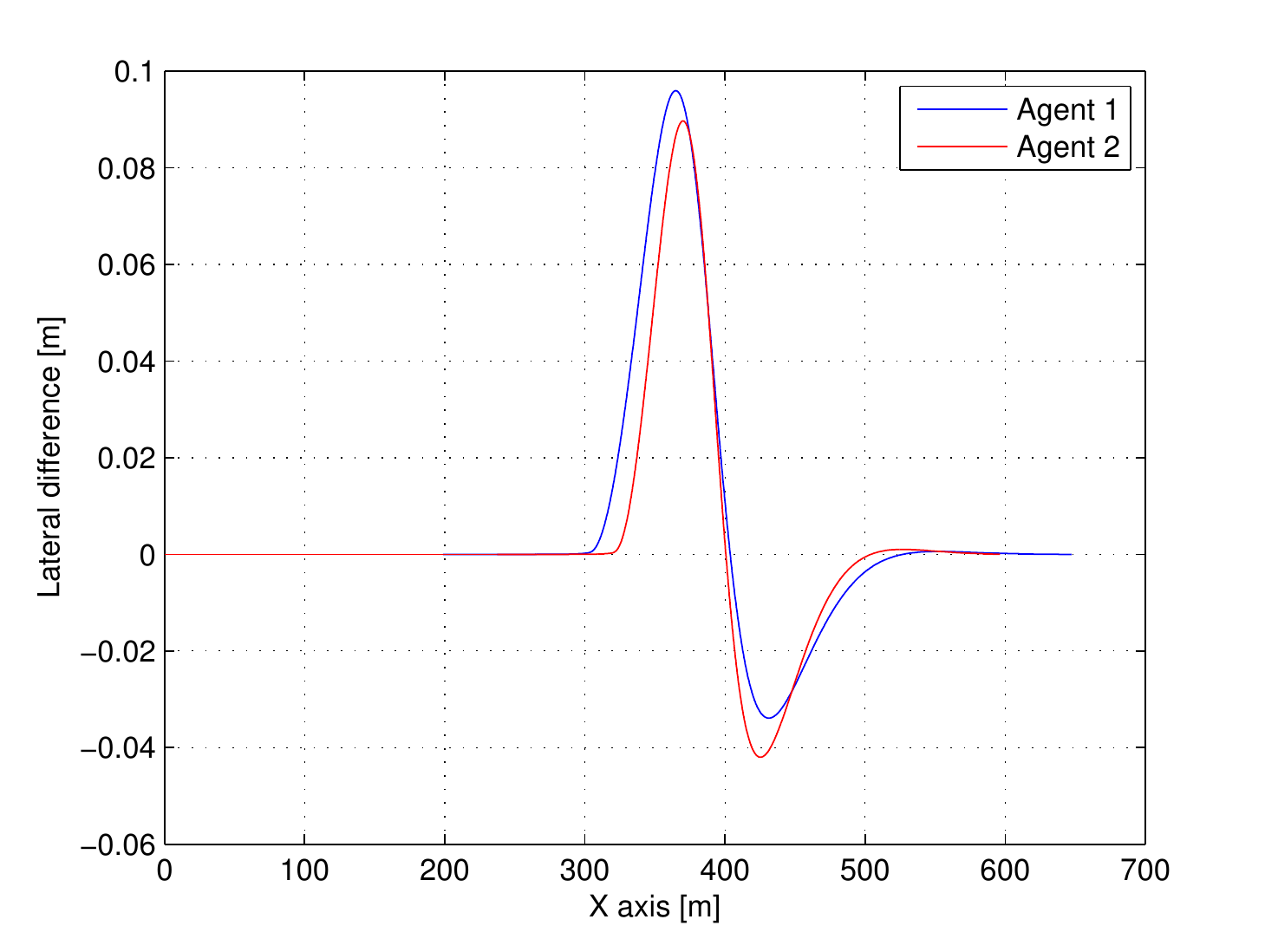}
	\caption{Lateral difference between the trajectories of the agents computed with the proposed decentralized algorithm and with a centralized one.}
	\label{Res1:1}
\end{figure}

Figures \ref{Res1:1} shows the lateral difference between the trajectories of the two agents computed with the proposed decentralized control algorithm and the centralized one. The maximum difference, between the trajectories computed with the centralized algorithm and the decentralized one, is $0.095m$ which is $5.59\%$ of the maximum lateral displacement. Thus, this proposed algorithm does not introduce further approximation with respect to a centralized continuation algorithm and it is able to compute a nearly optimal solution for the relaxed problem.

\begin{figure}[h!]
	\centering
	\includegraphics[width=1.0\columnwidth]{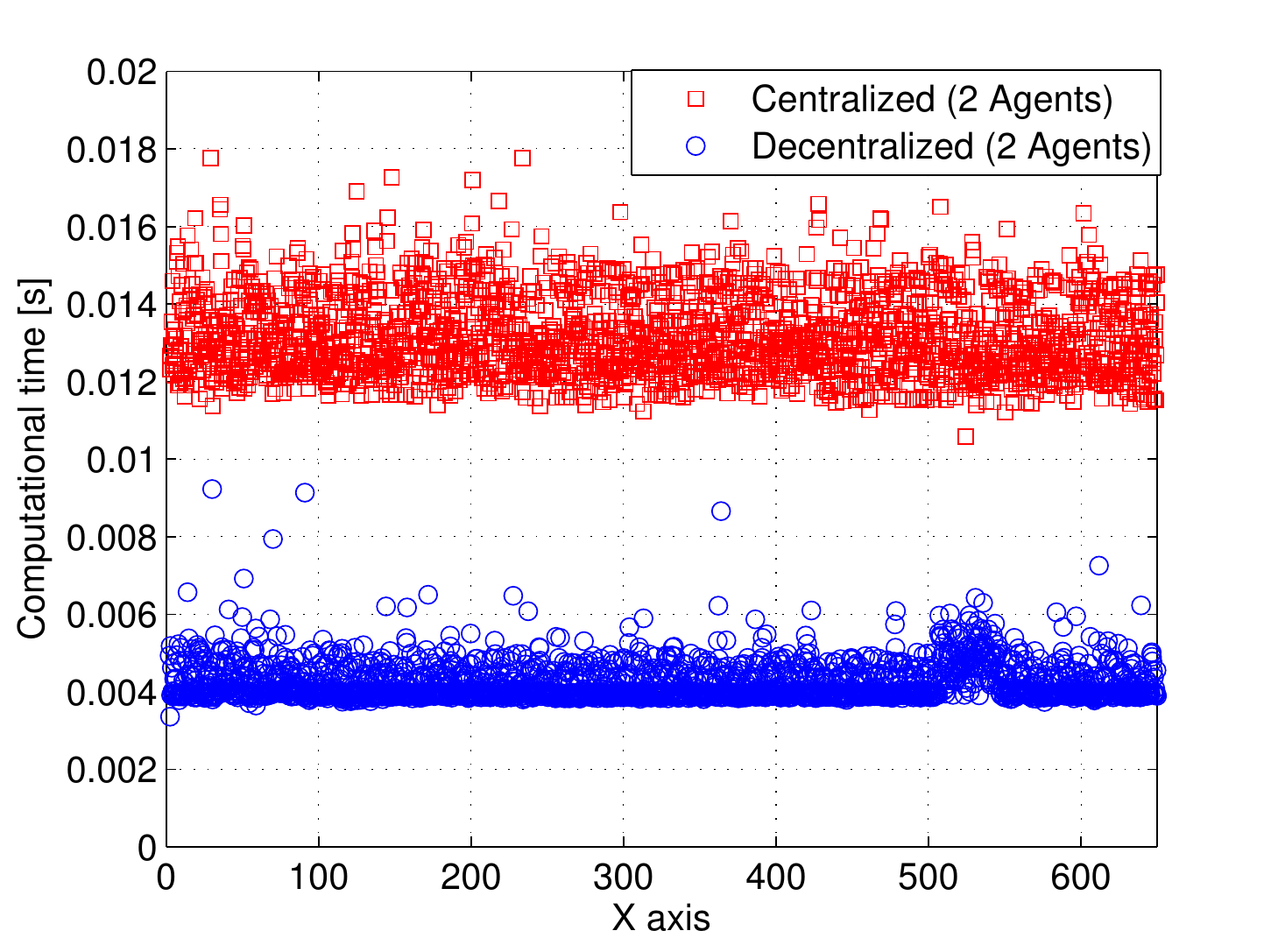}
	\caption{ Comparison between computational cost of the centralized and decentralized algorithms.}
	\label{Res1:2}
\end{figure}

Finally, it is important to analyze the computational cost. The centralized control strategy takes on average 1$2.9ms$ to compute the solution, while the decentralized one just $4.2ms$, as show in Figure \ref{Res1:2}.

\subsection{Communication method}
When the number of agents increases, a decentralized algorithm is necessary to limit the computational burden. In this section a simulation involving five agents is carried out and the computational time is analyzed. In this scenario, the proposed relaxed method to deal with inequality constraints (Eq. \ref{eq:16}) plays a crucial role. Coefficients and boundary conditions used in the simulation for the five agents can be found in Tables II, III and IV.

\begin{table}[h!]
	\centering
	\caption{Agents' initial conditions}
	\label{my-label}
	\begin{tabular}{lllllll}
		Agent           & $i=1$ & $i=2$ & $i=3$ & $i=4$ & $i=5$ & {[}units{]} \\ \hline
		$s_{i_{start}}$ & $2$   & $20$  & $50$  & $680$ & $480$ & {[}m{]}    
	\end{tabular}
\end{table}

\begin{figure}[h!]
	\centering
	\includegraphics[width=1.0\columnwidth]{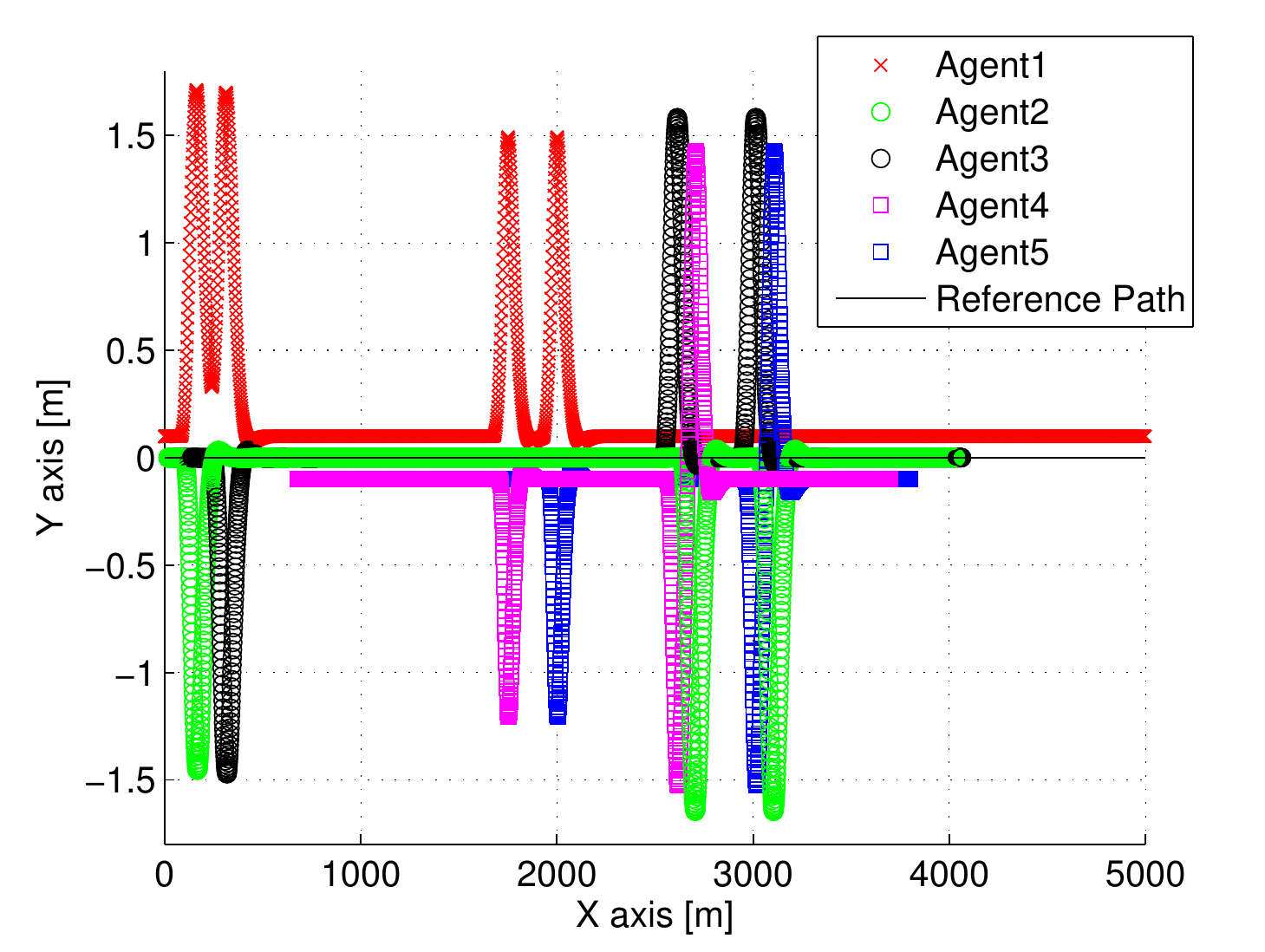}
	\caption{Trajectories of the five agents. Agent1 has the highest cruise velocity and it overtakes the others four agents. Agent2 and Agent3 have an intermediate cruise velocity and they overtake Agent4 and Agent5. Concluding, during the simulation there is a total of 8 overtaking maneuver.}
	\label{Res2:1}
\end{figure}

In Figure \ref{Res2:1} the trajectories of the five agents are shown. Agent1 travels at the highest cruise velocity and its starting position is the closest to the Y axis. Therefore, during the simulation it overtakes the slower agents that it encounters on the path.
	
\begin{figure}[h!]
	\centering
	\includegraphics[width=1.0\columnwidth]{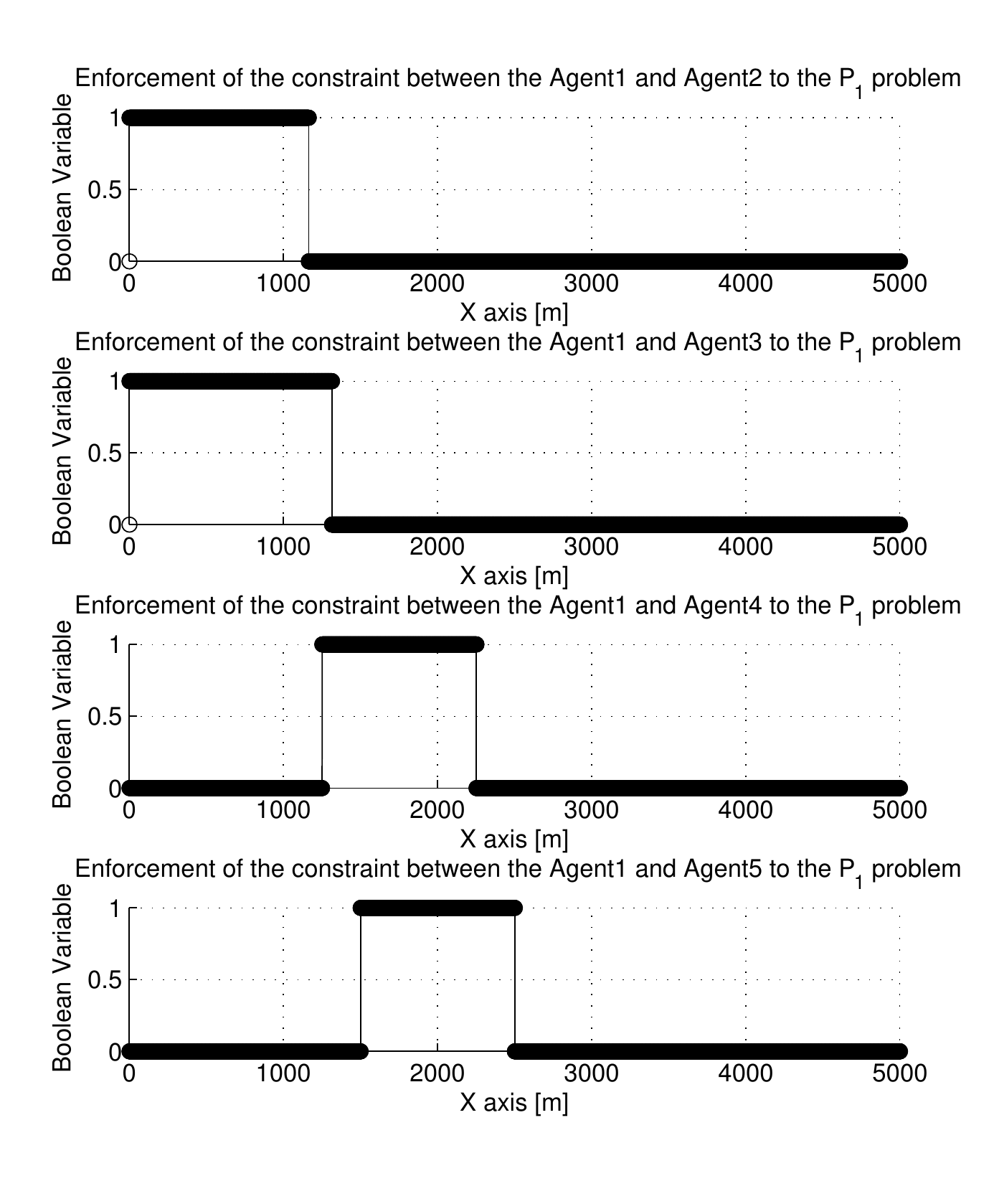}
	\caption{Coupling between Agent1 and the $i$-th agent in function of the traveled distance on the X axis. When the Boolean variable is set to one the Lagrange multiplier is different from zero.}
	\label{Res2:2}
\end{figure}

It is clear that the feasibility constraints, which couple Agent1 with the others, should be enforced to the $P_1$ just during the overtaking maneuvers. In Figure \ref{Res2:2}, a Boolean variable with values $1$ and $0$ is used to indicate, respectively, if the $i$-th constraint is enforced or is not enforced to $P_1$.

\begin{figure}[h!]
	\centering
	\includegraphics[width=1.0\columnwidth]{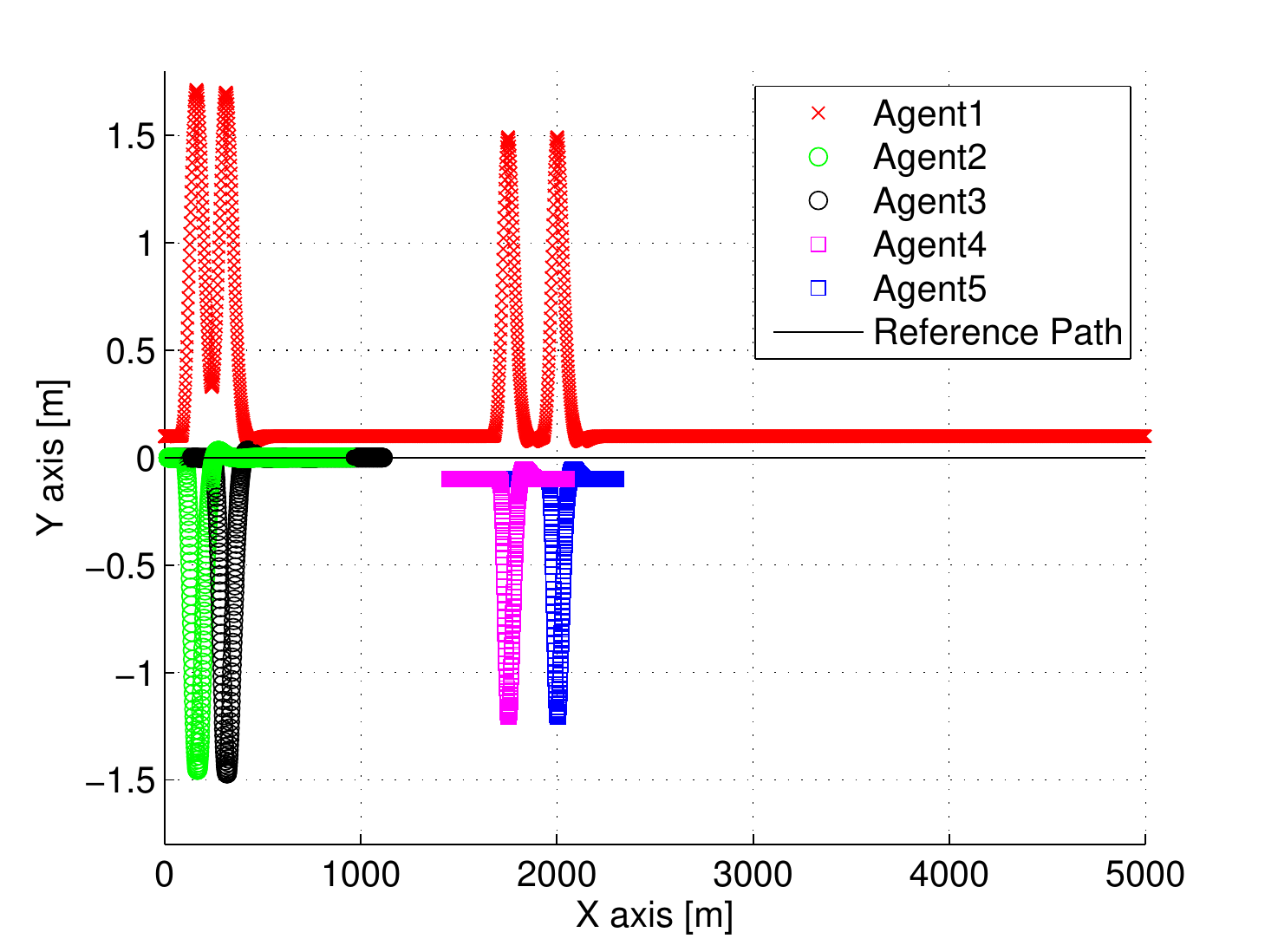}
	\caption{Trajectory of Agent1 with its prospective view. The trajectories of the others agents are reported when the related feasibility constraint are enforced to the $P_1$ problem, namely when the Boolean variable of Figure 9 is set to 1.}
	\label{Res2:3}
\end{figure}

Figure \ref{Res2:3} shows the trajectory of Agent1, and those of the other agents when the related feasibility constraints are enforced to $P_1$. Here it is possible to see that the relaxed constraints are correctly enforced to the problem just during the overtaking maneuvers.

\subsection{Complete Simulation}

\begin{figure}[h!]
	\centering
	\includegraphics[width=1.08\columnwidth]{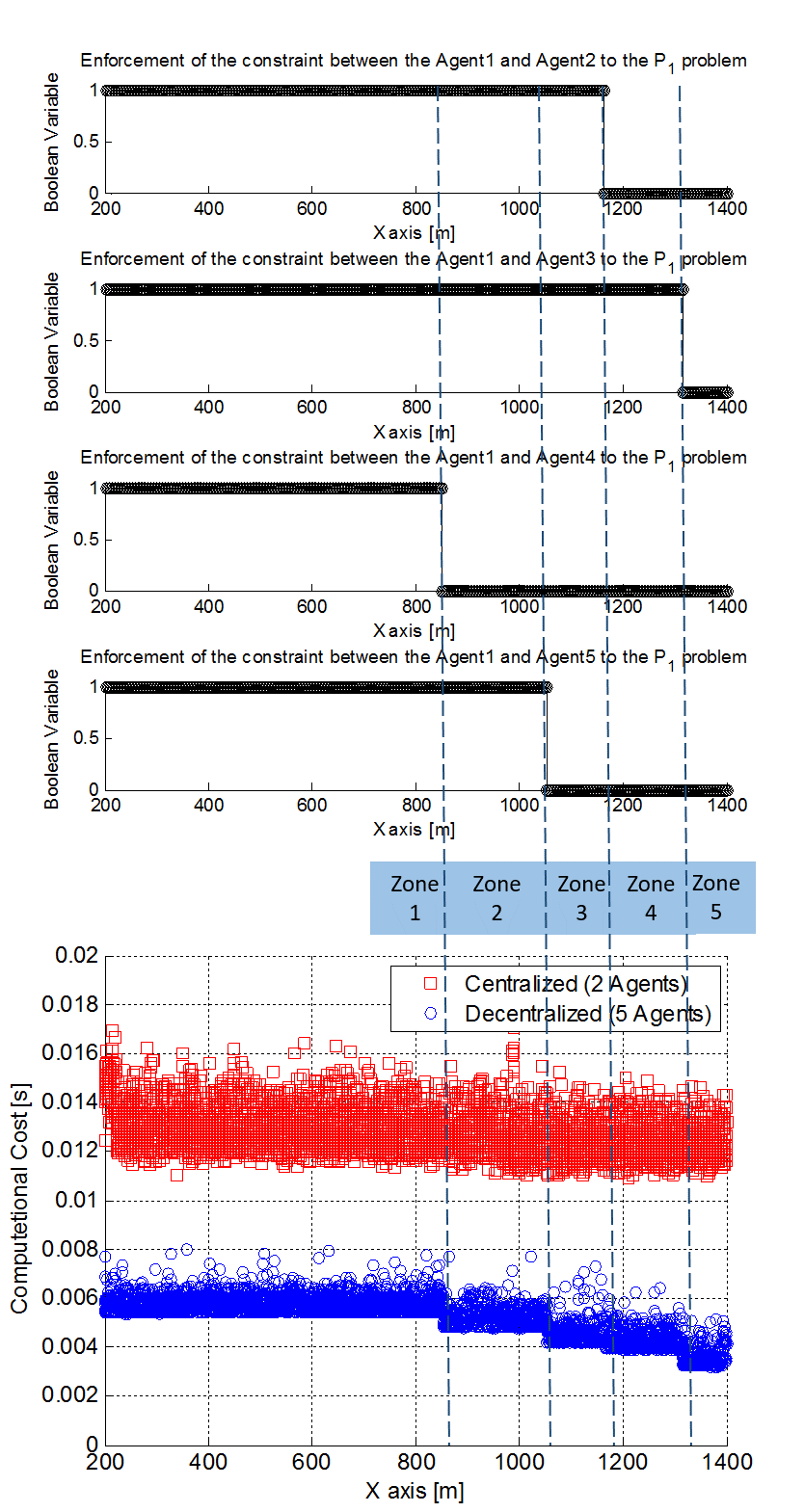}
	\caption{Trajectory of Agent1 with its prospective view. The trajectory of the others agents are reported when the related feasibility constraint are enforced to the $P_1$ problem, namely when the Boolean variable of Figure 10 is set to 1.}
	\label{Res3:1}
\end{figure}

Finally the algorithm is tested in the worst case scenario,
where all the feasibility constraints has to be enforced to the
$P_1$ problem (A video of the simulation can be found at {\small{ \url{http://youtu.be/wTfb5M1YH44} }}). Figure \ref{Res3:1} shows the
behavior of the computational cost as a function of the relaxed
constraints. In particular, Figure \ref{Res3:1} is divided in five zones,
numbered from $0$ to $4$, to indicate the number of enforced
constraints. The minimum computational cost, $3.85ms$, is
achieved when no constrains are enforced to the problem.
Furthermore, the maximum computational cost, $6.02ms$, is
reached when all the four constraints are enforced to the $P_1$ problem. Thus, the increment in computational cost, between the unconstrained problem and the one where all four constrains are enforced, is $2.17ms$. This increment is small when compared with the centralized approach which took $12.5ms$ to solve a problem involving two agents.

\section{Conclusions}
 In this paper a decentralized control algorithm for dynamically decoupled system, coupled by feasibility constraint, is presented. The algorithm, similarly to continuation methods, uses the derivative of the optimal solution to approximate the behavior of the system. This strategy allows to decouple and to parallelize computations.
 
Moreover, a relaxed approach to deal with inequalities constraints is introduced. This approach allows one to eliminate the discontinuity introduced by the KKT conditions; but it is able to recognize when an inequality constraint does not influence optimality and thus should not be enforced on the problem.

The algorithm has been successfully tested in simulation in a cooperative driving scenario. The control logic is able to compute a solution near the global optimal with a decentralized strategy. The size of the problem is reduced when the coupling between agents is not relevant, thus the computational burden in reduced. Finally, the computational cost of a simulation involving five coupled agents is compared with a centralized control problem involving two agents. This comparison underlines the advantage of the decentralized control strategy which took, on average, $50\%$ less time to solve the optimal control problem, though the dimension of the problem is four times larger.

\section{ACKNOWLEDGMENTS}
The authors thank Dr. Jacopo Guanetti from UC Berkeley from for his feedback on the manuscript presented.

%%%%%%%%%%%%%%%%%%%%%%%%%%%%%%%%%%%%%%%%%%%%%%%%%%%%%%%%%%%%%%%%%%%%%%%%%%%%%%%%
\bibliographystyle{IEEEtran} 
\bibliography{IEEEabrv,mybibfile}

\begin{thebibliography}{10}
\providecommand{\url}[1]{#1}
\csname url@samestyle\endcsname
\providecommand{\newblock}{\relax}
\providecommand{\bibinfo}[2]{#2}
\providecommand{\BIBentrySTDinterwordspacing}{\spaceskip=0pt\relax}
\providecommand{\BIBentryALTinterwordstretchfactor}{4}
\providecommand{\BIBentryALTinterwordspacing}{\spaceskip=\fontdimen2\font plus
\BIBentryALTinterwordstretchfactor\fontdimen3\font minus
  \fontdimen4\font\relax}
\providecommand{\BIBforeignlanguage}[2]{{%
\expandafter\ifx\csname l@#1\endcsname\relax
\typeout{** WARNING: IEEEtran.bst: No hyphenation pattern has been}%
\typeout{** loaded for the language `#1'. Using the pattern for}%
\typeout{** the default language instead.}%
\else
\language=\csname l@#1\endcsname
\fi
#2}}
\providecommand{\BIBdecl}{\relax}
\BIBdecl

\bibitem{1}
A.~Bemporad and D.~Barcelli, ``Decentralized model predictive control,'' in
  \emph{Networked control systems}.\hskip 1em plus 0.5em minus 0.4em\relax
  Springer, 2010, pp. 149--178.

\bibitem{2}
A.~Vahidi and W.~Greenwell, ``A decentralized model predictive control approach
  to power management of a fuel cell-ultracapacitor hybrid,'' in \emph{American
  Control Conference, 2007. ACC'07}.\hskip 1em plus 0.5em minus 0.4em\relax
  IEEE, 2007, pp. 5431--5437.

\bibitem{3}
A.~Mohammadi and M.~B. Menhaj, ``Formation control and obstacle avoidance for
  nonholonomic robots using decentralized mpc,'' in \emph{Networking, Sensing
  and Control (ICNSC), 2013 10th IEEE International Conference on}.\hskip 1em
  plus 0.5em minus 0.4em\relax IEEE, 2013, pp. 112--117.

\bibitem{4}
Z.~Ma, I.~Hiskens, and D.~Callaway, ``A decentralized mpc strategy for charging
  large populations of plug-in electric vehicles,'' \emph{IFAC Proceedings
  Volumes}, vol.~44, no.~1, pp. 10\,493--10\,498, 2011.

\bibitem{5}
S.~Devasia, D.~Iamratanakul, G.~Chatterji, and G.~Meyer, ``Decoupled
  conflict-resolution procedures for decentralized air traffic control,''
  \emph{IEEE Transactions on Intelligent Transportation Systems}, vol.~12,
  no.~2, pp. 422--437, 2011.

\bibitem{6}
V.~R. Desaraju and J.~P. How, ``Decentralized path planning for multi-agent
  teams in complex environments using rapidly-exploring random trees,'' in
  \emph{Robotics and Automation (ICRA), 2011 IEEE International Conference
  on}.\hskip 1em plus 0.5em minus 0.4em\relax IEEE, 2011, pp. 4956--4961.

\bibitem{7}
D.~Raimondo, L.~Magni, and R.~Scattolini, ``Decentralized mpc of nonlinear
  systems: An input-to-state stability approach,'' \emph{International Journal
  of Robust and Nonlinear Control}, vol.~17, no.~17, pp. 1651--1667, 2007.

\bibitem{8}
A.~Richards and J.~How, ``Decentralized model predictive control of cooperating
  uavs,'' in \emph{Decision and Control, 2004. CDC. 43rd IEEE Conference on},
  vol.~4.\hskip 1em plus 0.5em minus 0.4em\relax IEEE, 2004, pp. 4286--4291.

\bibitem{9}
T.~Keviczky, F.~Borrelli, K.~Fregene, D.~Godbole, and G.~J. Balas,
  ``Decentralized receding horizon control and coordination of autonomous
  vehicle formations,'' \emph{IEEE Transactions on Control Systems Technology},
  vol.~16, no.~1, pp. 19--33, 2008.

\bibitem{trodden2013cooperative}
P.~Trodden and A.~Richards, ``Cooperative distributed mpc of linear systems
  with coupled constraints,'' \emph{Automatica}, vol.~49, no.~2, pp. 479--487,
  2013.

\bibitem{10}
P.~Ogren, M.~Egerstedt, and X.~Hu, ``A control lyapunov function approach to
  multi-agent coordination,'' in \emph{Decision and Control, 2001. Proceedings
  of the 40th IEEE Conference on}, vol.~2.\hskip 1em plus 0.5em minus
  0.4em\relax IEEE, 2001, pp. 1150--1155.

\bibitem{11}
A.~Bicchi and L.~Pallottino, ``On optimal cooperative conflict resolution for
  air traffic management systems,'' \emph{IEEE Transactions on Intelligent
  Transportation Systems}, vol.~1, no.~4, pp. 221--231, 2000.

\bibitem{trodden2010distributed}
P.~Trodden and A.~Richards, ``Distributed model predictive control of linear
  systems with persistent disturbances,'' \emph{International Journal of
  Control}, vol.~83, no.~8, pp. 1653--1663, 2010.

\bibitem{12}
S.~L. Richter and R.~A. Decarlo, ``Continuation methods: Theory and
  applications,'' \emph{IEEE Transactions on Systems, Man, and Cybernetics},
  no.~4, pp. 459--464, 1983.

\bibitem{13}
T.~Ohtsuka, ``Continuation/gmres method for fast algorithm of nonlinear
  receding horizon control,'' in \emph{Decision and Control, 2000. Proceedings
  of the 39th IEEE Conference on}, vol.~1.\hskip 1em plus 0.5em minus
  0.4em\relax IEEE, 2000, pp. 766--771.

\bibitem{14}
------, ``A continuation/gmres method for fast computation of nonlinear
  receding horizon control,'' \emph{Automatica}, vol.~40, no.~4, pp. 563--574,
  2004.

\bibitem{15}
A.~E. Bryson, \emph{Applied optimal control: optimization, estimation and
  control}.\hskip 1em plus 0.5em minus 0.4em\relax CRC Press, 1975.

\bibitem{16}
D.~Liberzon, \emph{Calculus of variations and optimal control theory: a concise
  introduction}.\hskip 1em plus 0.5em minus 0.4em\relax Princeton University
  Press, 2012.

\bibitem{17}
C.~Kelley, ``Iterative methods for linear and nonlinear equations, siam,
  philadelphia, 1995,'' \emph{MR 96d}, vol. 65002.

\bibitem{18}
Y.~Saad and M.~H. Schultz, ``Gmres: A generalized minimal residual algorithm
  for solving nonsymmetric linear systems,'' \emph{SIAM Journal on scientific
  and statistical computing}, vol.~7, no.~3, pp. 856--869, 1986.

\bibitem{19}
T.~Hashimoto, Y.~Yoshioka, and T.~Ohtsuka, ``Receding horizon control with
  numerical solution for spatiotemporal dynamic systems,'' in \emph{Decision
  and Control (CDC), 2012 IEEE 51st Annual Conference on}.\hskip 1em plus 0.5em
  minus 0.4em\relax IEEE, 2012, pp. 2920--2925.

\bibitem{21}
D.~A. Knoll and D.~E. Keyes, ``Jacobian-free newton--krylov methods: a survey
  of approaches and applications,'' \emph{Journal of Computational Physics},
  vol. 193, no.~2, pp. 357--397, 2004.

\bibitem{22}
T.~Hashimoto, Y.~Yoshioka, and T.~Ohtsuka, ``Receding horizon control for
  nonlinear parabolic partial differential equations with boundary control
  inputs,'' in \emph{Decision and Control (CDC), 2010 49th IEEE Conference
  on}.\hskip 1em plus 0.5em minus 0.4em\relax IEEE, 2010, pp. 6920--6925.

\bibitem{23}
T.~Ohtsuka and H.~A. Fujii, ``Real-time optimization algorithm for nonlinear
  receding-horizon control,'' \emph{Automatica}, vol.~33, no.~6, pp.
  1147--1154, 1997.

\bibitem{24}
A.~Gray, Y.~Gao, T.~Lin, J.~K. Hedrick, H.~E. Tseng, and F.~Borrelli,
  ``Predictive control for agile semi-autonomous ground vehicles using motion
  primitives,'' in \emph{American Control Conference (ACC), 2012}.\hskip 1em
  plus 0.5em minus 0.4em\relax IEEE, 2012, pp. 4239--4244.

\bibitem{25}
Y.~Gao, T.~Lin, F.~Borrelli, E.~Tseng, and D.~Hrovat, ``Predictive control of
  autonomous ground vehicles with obstacle avoidance on slippery roads,'' in
  \emph{ASME 2010 dynamic systems and control conference}.\hskip 1em plus 0.5em
  minus 0.4em\relax American Society of Mechanical Engineers, 2010, pp.
  265--272.

\bibitem{26}
P.~Falcone, F.~Borrelli, J.~Asgari, H.~Tseng, and D.~Hrovat, ``Low complexity
  mpc schemes for integrated vehicle dynamics control problems,'' in \emph{9th
  international symposium on advanced vehicle control}, 2008.

\bibitem{27}
U.~Rosolia, F.~Braghin, A.~Alleyne, and E.~Sabbioni, ``Nlmpc for real time path
  following and collision avoidance,'' \emph{SAE International Journal of
  Passenger Cars-Electronic and Electrical Systems}, vol.~8, no. 2015-01-0313,
  pp. 401--405, 2015.

\bibitem{28}
U.~Rosolia, S.~De~Bruyne, and A.~G. Alleyne, ``Autonomous vehicle control: A
  nonconvex approach for obstacle avoidance,'' \emph{IEEE Transactions on
  Control Systems Technology}, 2016.

\bibitem{29}
A.~Micaelli and C.~Samson, ``Trajectory tracking for unicycle-type and
  two-steering-wheels mobile robots,'' Ph.D. dissertation, INRIA, 1993.

\end{thebibliography}

\end{document}